\documentclass{ifacconf}

\usepackage{graphicx,epstopdf}
\usepackage{natbib}        
\usepackage{amsmath}
\usepackage{amssymb}
\usepackage{siunitx}

\usepackage{listings}
\usepackage{tikz}
\usepackage{tikzscale}

\usepackage[utf8]{inputenc}
\usepackage{pgfplots} 
\usepackage{pgfgantt}
\usepackage{pdflscape}
\pgfplotsset{compat=newest} 
\pgfplotsset{plot coordinates/math parser=false}

\usetikzlibrary{shapes,arrows,patterns}

\definecolor{codegray}{rgb}{0.5,0.5,0.5}
\definecolor{darkRed}{rgb}{0.7,0,0}
\definecolor{darkGreen}{rgb}{0,0.5,0}
\definecolor{darkBlue}{rgb}{0,0,0.6}
\definecolor{Black}{rgb}{0,0,0}
\definecolor{Grey}{rgb}{0.7,0.7,0.7}
\definecolor{Orange}{rgb}{0.8,0.4,0}
\definecolor{darkCyan}{rgb}{0,0.6,0.6}
\definecolor{Purple}{rgb}{0.5,0.0,0.5}
\definecolor{Green}{rgb}{0.6,0,0.3}
\definecolor{Gold}{rgb}{0.7,0.7,0}
\definecolor{lightRedPink}{rgb}{0.7765,0.6314,0.7216}
\definecolor{lightGreen}{rgb}{0.0,0.7,0.4}
\definecolor{lightBlue}{rgb}{0.6,0.6,1.0}

\DeclareRobustCommand\DarkRedLine{\tikz[baseline=-0.6ex]\draw[color=darkRed, thick, line width=2pt] (0,0)--(0.5,0);}
\DeclareRobustCommand\DarkBlueLine{\tikz[baseline=-0.6ex]\draw[color=darkBlue, thick, line width=2pt] (0,0)--(0.5,0);}

\DeclareRobustCommand\BlackLine{\tikz[baseline=-0.6ex]\draw[color=Black, thick, line width=2pt] (0,0)--(0.5,0);}
\DeclareRobustCommand\GreyLine{\tikz[baseline=-0.6ex]\draw[color=Grey, thick, line width=2pt] (0,0)--(0.5,0);}

\DeclareRobustCommand\lightRedPinkLine{\tikz[baseline=-0.6ex]\draw[color=lightRedPink, thick, line width=2pt] (0,0)--(0.5,0);}

\DeclareRobustCommand\lightBlueLine{\tikz[baseline=-0.6ex]\draw[color=lightBlue, thick, line width=2pt] (0,0)--(0.5,0);}

\newcommand{\darkGreenCircle}{%
   \begin{tikzpicture}[baseline={(0,0.03)}]
      \fill[darkGreen] (0.15,0.12) circle [radius=0.1];
   \end{tikzpicture}%
}

\DeclareRobustCommand{\purpleCircle}{%
   \begin{tikzpicture}[baseline={(0,0.03)}] 
      \draw[-,Purple,solid,line width=0.9pt] (0.15,0.12) circle [radius=0.1];
   \end{tikzpicture}%
}

\DeclareRobustCommand{\darkCyanCircle}{%
   \begin{tikzpicture}[baseline={(0,0.03)}] 
      \draw[-,darkCyan,solid,line width=0.9pt] (0.15,0.12) circle [radius=0.1];
   \end{tikzpicture}%
}

\DeclareRobustCommand{\goldCircle}{%
   \begin{tikzpicture}[baseline={(0,0.03)}] 
      \draw[-,Gold,solid,line width=0.9pt] (0.15,0.12) circle [radius=0.1];
   \end{tikzpicture}%
}

\usepackage{eso-pic}
\AddToShipoutPictureBG*{%
\AtPageUpperLeft{%
\setlength\unitlength{1in}%
\hspace*{\dimexpr0.5\paperwidth\relax}
\makebox(0,-1.75)[c]{
\begin{tabular}{c c}
Mathyn van Dael \emph{et al.},
GraFIT: A toolbox for fast and accurate frequency response identification \\
in Gravitational Wave Detectors, 
uploaded to ArXiv \today \\
\end{tabular}}}}

\begin{document}
\begin{frontmatter}


\thanks[footnoteinfo]{This work has been funded by the Netherlands Organisation for Scientific Research (NWO) under grant number 680.92.18.02. }

\title{GraFIT: A toolbox for fast and accurate frequency response identification in Gravitational Wave Detectors}

\author{Mathyn van Dael\textsuperscript{1,3}, Max van Haren\textsuperscript{1}, Gert Witvoet\textsuperscript{1,2},} 
\author{Bas Swinkels\textsuperscript{3} and Tom Oomen\textsuperscript{1,4}}

\address{1 \textit{Eindhoven University of Technology, dept. of Mechanical Engineering, Control Systems Technology} \\ 
\textit{Eindhoven, The Netherlands, email: m.r.v.dael@tue.nl} \\
2 TNO, Optomechatronics Department, Delft, The Netherlands\\
3 Nikhef, Amsterdam, The Netherlands \\
4 Delft Center for Systems and Control, Delft University of Technology, Delft, The Netherlands}

\begin{abstract}                
Frequency response function (FRF) measurements are widely used in Gravitational Wave (GW) detectors, e.g., for the design of controllers, calibrating signals and diagnostic problems with system dynamics. The aim of this paper is to present GraFIT: a toolbox that enables fast, inexpensive, and accurate identification of FRF measurements for GW detectors compared to the commonly used approaches, including common spectral analysis techniques. The toolbox consists of a single function to estimate the frequency response function for both open-loop and closed-loop systems and for arbitrary input and output dimensions. The toolbox is validated on two experimental case studies of the Virgo detector, illustrating more than a factor 3 reduction in standard deviation of the estimate for the same measurement times, and comparable standard deviations with up to 10 times less data for the new method with respect to the currently implemented Spectral Analysis method.
\end{abstract}


\end{frontmatter}


\section{Introduction}

Frequency domain models are essential in Gravitational Wave (GW) detectors for a variety of purposes, e.g, control design, calibration of signals and diagnosing problems with system dynamics. Frequency Reponse Functions (FRF) measurements are commonly used to represent system dynamics in the frequency domain because they are accurate, user-friendly and inexpensive to obtain \Citep{pintelon2012system}. Identifying an FRF requires the user to choose the perturbation signal and the identification method. Non-periodic perturbation signals (typically filtered white noise) combined with the Spectral Analysis (SA) \Citep{Bendat_SA,pintelon2012system} method have been the baseline for FRF identification in almost all application domains, including the GW community. The SA method is typically accurate, but it requires a sufficient number of data segments to average over to obtain an accurate estimate, at the expense of the frequency resolution of the estimation. Essentially, if sufficient measurement data is taken, sufficient accuracy can be obtained.

In more recent literature, new methods for FRF identification have been proposed \Citep{Schoukens2005,Hagg2016,Lataire2016}, among which the Local Polynomial Method (LPM) \Citep{Schoukens2009}. LPM has been developed to better address transient errors, resulting from dynamic transients or leakage effects, by leveraging the observation that such errors exhibit smooth frequency characteristics \Citep[Appendix 6.B]{pintelon2012system}. By explicitly estimating this transient effect, LPM effectively suppresses it in the output signal. LPM models the transient in the frequency domain using a polynomial function, a method later extended to rational models in the Local Rational Method (LRM) \Citep{McKelvey2012}. Compared to the SA method which uses windowing \Citep{Schoukens2006}, LPM has shown significant improvements in reducing leakage effects \Citep{Gevers2011}. LPM has furthermore been shown to be very effective for systems with large dynamic transients such as thermal systems \Citep{Evers2020_ThermalIdentification} and LRM for mechanical systems with lightly damped resonant dynamics \Citep{VOORHOEVE2018129}. Finally, the local modelling approach has been shown to be very data-efficient \Citep{VOORHOEVE2018129,Tacx2024}.

Although the identification of FRFs are standard in GW, the main goal of this paper is to obtain much more data-efficient FRF models that have a better accuracy versus data size ratio. In this paper, the Gravitational Waves Advanced Frequency Response Identification Toolbox (GraFIT) is therefore presented, which uses the LPM/LRM method to identify FRFs for GW detectors. The toolbox is validated on two experimental case studies of the Virgo detector and the toolbox is freely available\footnote{Toolbox available at https://github.com/MathynVanD/GraFIT.git}. The toolbox is designed to be user-friendly and to be directly usable in the GW community, aiming to implement developments from the control community and to make them accessable in the GW community.

The paper is organised as follows. In Section \ref{sec:ApplicationSetting}, the application setting is discussed to highlight the importance of FRF identification in the operation of GW detectors. In Section \ref{sec:LocalModellingToolbox}, the theory behind the approach as well as the toolbox itself are presented. In Section \ref{sec:ExperimentalCaseStudies}, two experimental case studies are presented to illustrate the performance improvement of the local modelling approach with respect to the standard SA method and finally in Section \ref{sec:Conclusions} conclusions on the work are given.

\section{Application domains}
\label{sec:ApplicationSetting}

GW detectors such as Advanced Virgo+ (AdV+) \Citep{Acernese_Virgo_Overview} are kilometer-scale interferometers consisting of a vast number of optics to detect indulations in the arm lengths in the order of \SI{1e-18}{\metre}. FRF identification plays an important role in various aspects of GW detectors; three of such application domains are discussed next.

\subsection{Suspension systems for optics}
\label{sec:appDomain_SuspensionSystems}

Suspension systems such as the ones in AdV+ \citep{braccini_SuperAttenuator_Perf_Virgo, Heijningen2019} are critical to minimizing the motion of the optics due to ground motion. These suspensions use low-stiffness isolators to passively isolate ground motion, reducing for example the mirror motion by over 15 orders of magnitude above \SI{10}{\hertz} \citep{braccini_SuperAttenuator_Perf_Virgo}. To damp the typically lowly-damped modes and thus to minimize the Root-Mean-Square (RMS) motion, feedback controllers are used. A FRF of the system is typically sufficient for the design of these controllers. Identification methods for obtaining these FRFs are typically preferred over modelling tools since they are fast and inexpensive to obtain and less prone to errors compared to modelling methods. An accurate estimation of the FRF is however essential for the performance of the controller.

\subsection{Relative mirror position control}
\label{sec:appDomain_RelativeMirrorPositionControl}

Feedback control for the relative distances between the optics is essential to achieving the required sensitivity of the detector \Citep{Bersanetti_DRMI_ErrorSignals}. Obtaining models for the control design by accurately simulating the system dynamics is difficult since imperfections in the optics are difficult to model and the system exhibits time-varying behaviour \Citep{vanDael2024_OnlineDecoupling}. Instead, the FRF is measured by having a dedicated experiment where the optics are perturbed and the response is measured. Since this experiment requires downtime of the machine, an identification method that requires less data while maintaining similair accuracy to the classical identification method is preferred to minimize the downtime of the detector.

\subsection{Noise budget}

Noise budgets are used to identify the limiting disturbances at each frequency for the detector sensitivity \Citep{Bersanetti_AdV_Status,Buikema_aLIGO_StatusO3}. To determine the contribution of a disturbance, both a model of the disturbance and the FRF of the coupling to the sensitivity is required. Identifying the FRF is often preferred over modelling due to the complexity of the system and the time-varying behaviour. The quality of the noise projection heavily depends on the quality of the measured FRF, thus requiring accurate FRFs.

\section{FRF Identification using GraFIT}
\label{sec:LocalModellingToolbox}

This section presents the method behind GraFIT to obtain non-parametric models. First, the identification setting is discussed, after which the LRM method in GraFIT for non-parametric identification is presented and finally, the toolbox and its use are illustrated.

\subsection{Identification problem}
\label{sec:IdentificationProblem}

\begin{figure}[bt]
   \centering
   \resizebox{0.6\columnwidth}{!}{
           \includegraphics[]{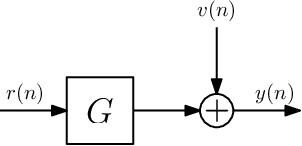}}
   \caption{Standard open-loop identification problem where the goal is to identify $G$ using the input signal $r(n)$ and noisy output signal $y(n)$ which is perturbed a disturbance $v(n)$. }
   \label{fig:BlockDiagram_OL_Identification}
\end{figure}

Consider the basic identification problem as shown in Fig. \ref{fig:BlockDiagram_OL_Identification}. The objective is to obtain the FRF of the dynamical system $G$ using the input signal $r(n)$ and the output signal $y(n)$, which is perturbed by the unknown disturbance $v(n)$ and with $n = 0,\: 1,\: ... ,\: N-1$, and $N$ the number of samples.

There are different possibilities for perturbation signals but here the standard approach using filtered white noise is considered as external perturbation for $r(n)$. The noisy output $y(n)$, perturbated by some noise $v(n)$, is measured and the DFT of the input and output signals are computed through

\begin{equation}
   X(k) = \frac{1}{\sqrt{N}}\sum_{n=0}^{N-1} x(n)e^{j\omega_k n}, 
   \label{eq:DFT_Equation}
\end{equation}

with $x=r,y,v$ and $X(k)=R(k),Y(k),V(k)$ and 

\begin{equation}
   \omega_k = \frac{2\pi k}{N},
\end{equation}

where $\omega_k$ represents the frequency grid and and $k$ relates to the frequency bin $\omega_k$. The following input-output relation is then obtained

\begin{equation}
   Y(k) = G(\Omega_k)R(k) + T(\Omega_k) + V(k).
   \label{eq:SISO_OL_InputOutRelation}
\end{equation}

where $\Omega_k=e^{-j\omega_k n}$ represents the generalized frequency variable. The additional term $T(\Omega_k)$ represents the transient effect of both the system dynamics $G$ as well as the noise term when filtered white noise is used. The goal is then to have an identification procedure that uses as little data as possible to identify the FRF $G(\Omega_k)$, which requires effectively handling both the transient effect $T(\Omega_k)$ and noise term $V(k)$.

\subsection{FRF identification using local models}
\label{sec:FRF_using_LocalModels}

The local modelling method, first presented in \citep{Schoukens2009} for polynomial models and later extended to rational models in \citep{McKelvey2012}, tackles these downsides by simultaneously estimating and suppressing the transient term $T(\Omega_k)$ next to estimating the system dynamics $G(\Omega_k)$. The estimated transient term accounts for any smooth behaviour in the frequency domain not resulting from the system dynamics, thus capturing both physical transient effects as well as leakage errors. The local modelling method has shown to significantly reduce the effect of leakage compared to the SA method using windowing \Citep{Gevers2011} as well as be effective in fast and accurate identification of systems with large dynamic transients such as thermal systems \Citep{Evers2020_ThermalIdentification} or mechanical systems with lightly damped resonant dynamics \Citep{VOORHOEVE2018129}.

The key concept is to estimate a rational model for both the system dynamics $G(\Omega_k)$ as well as the transient term $T(\Omega_k)$ in a local frequency window $l \in \mathbb{Z}_{\left[-W,\: W \right]}$ with $2W+1$ the window size around the frequency bin $\Omega_k$. The estimated output DFT at the frequency bin $k+l$ is then given by

\begin{equation}
   \widehat{Y}(k+l) = \widehat{G}(\Omega_{k+l})R(k+l) + \widehat{T}(\Omega_{k+l}).
\end{equation}

Both $\widehat{G}(\Omega_{k+l})$ and $\widehat{T}(\Omega_{k+l})$ are parametrized as rational functions of frequency. The system dynamics is parametrized as

\begin{equation}
   \widehat{G}(\Omega_{k+l}) = \frac{A_{k+l}}{D_{k+l}},
\end{equation}

where both $A_{k+l}$ and $D_{k+l}$ are polynomials in the frequency given by

\begin{equation}
      A_{k+l} = G(\Omega_k)+\sum_{i=1}^{L_a} a_i(k)l^i,
\end{equation}

and

\begin{equation}
      D_{k+l} = 1 + \sum_{i=1}^{L_d} d_i(k)l^i.
\end{equation}

Similairly, the transient term is parametrized as

\begin{equation}
   \widehat{T}(\Omega_{k+l}) = \frac{B_{k+l}}{D_{k+l}},
\end{equation}

with

\begin{equation}
      B_{k+l} = T(\Omega_k)+\sum_{i=1}^{L_b} b_i(k)l^i.
\end{equation}

Note that $\widehat{G}(\Omega_{k+l})$ and $\widehat{T}(\Omega_{k+l})$ have the same demoninator since they share the same poles \Citep{McKelvey2012}. 

The estimation of the local model can be defined as a Least Squares (LS) problem. First, the optimization variables, which are the model coefficients, are gathered in the parameter vector $\Theta(k)\in \mathbb{C}^{n_{\Theta}}$ with $n_{\Theta}=L_a+L_b+n_y L_d+2$ the number of parameters in the local model and $n_y$ the number of outputs in the estimation procedure, i.e.,

\begin{equation}
   \begin{split}
      \Theta(k) &= \left[\Theta_A(k) \:\:\: \Theta_B(k) \:\:\: \Theta_D(k)\right], \\
      \Theta_A(k) &= \left[ G(\Omega_k) \:\:\: a_1(k) \:\:\: \dots \:\:\: a_{L_a}(k) \right], \\
      \Theta_B(k) &= \left[ T(\Omega_k) \:\:\: b_1(k) \:\:\: \dots \:\:\: b_{L_b}(k) \right], \\
      \Theta_D(k) &= \left[ d_1(k) \:\:\: \dots \:\:\: d_{L_d}(k) \right].
   \end{split}
\end{equation}

The find the parameters $\Theta(k)$, the squared sum of the error between the measured output $Y(k+l)$ and the modelled output $\widehat{Y}(k+l)$ is minimized, i.e.,

\begin{equation}
   \widehat{\Theta}(k) = \mathrm{arg} \min_{\Theta(k)} \sum_{l=-W}^{W} \left| Y(k+l) - \widehat{Y}\left(k+l,\: \Theta(k)\right) \right|^2. 
   \label{eq:LS_Problem_Rational_Standard}
\end{equation}

Note that in the current formulation of $\widehat{Y}\left(k+l,\: \Theta(k)\right)$ the parametrization is not linear in $\Theta$. While there are several approaches to solving this LS problem, see e.g. \Citep{VOORHOEVE2018129}, the approach used here is the most straightforward and typically leads to satisfactory results. By weighting the cost function with the denominator $D_{k+l}$, the LS problem is rewritten into the linear LS problem

\begin{equation}
   \widehat{\Theta}(k) = \mathrm{arg} \min_{\Theta(k)} \sum_{l=-W}^{W} \left| Y(k+l) - \Theta(k) K(k+l)\right|^2.
   \label{eq:LS_Problem_Rational_Linear}
\end{equation}

with 

\begin{equation} 
   K(k+l) = 
   \begin{bmatrix}
      K_{a}(l)& \otimes \: U(k+l) \\
      K_{b}(l)& \\
      -K_{d}(l)& \otimes \: Y(k+l)
   \end{bmatrix} 
\end{equation}

and 

\begin{equation}
   \begin{split}
      K_{a}(l) &= \left[1\: l \: ... \: l^{L_a}\right]^T \\
      K_{b}(l) &= \left[1\: l \: ... \: l^{L_b}\right]^T \\
      K_{d}(l) &= \left[l \: ... \: l^{L_d}\right]^T.
   \end{split}
\end{equation}

The solution to the linear LS in \eqref{eq:LS_Problem_Rational_Linear} is then given by

\begin{equation}
   \widehat{\Theta}(k) = Y_{W}(k)K_{W}^H(k)\left(K_{W}(k)K_{W}^H(k)\right)^{-1}
\end{equation}

with $Y_{W}= \left[Y(k-W)\: ... \: Y(k+W)\right]^T$ and $K_{W}(k) = \left[K(k-W)\: ... \: K(k+W)\right]$. 

This LS problem is solved for every frequency bin $k$ to obtain the FRF of $G(\Omega_k)$. 


\subsection{Indirect identification}
\label{sec:IndirectIdentification}

\begin{figure}[bt]
   \centering
   \resizebox{0.8\columnwidth}{!}{
           \includegraphics[]{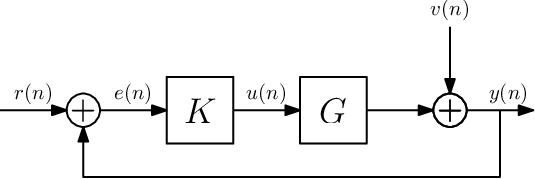}}
   \caption{Standard closed-loop identification problem, where the goal is to identify $G$ using the input signal $r(n)$ and noisy output signals $y(n)$ and $u(n)$.}
   \label{fig:BlockDiagram_CL_Identification}
\end{figure}

For systems operating in closed-loop, direct identification leads to biased FRFs, so this section addresses the identification procedure for systems operating in closed-loop. In Fig. \ref{fig:BlockDiagram_CL_Identification}, the block diagram for a system operating in closed-loop is shown. Here, the goal is to obtain the FRF of $G(\Omega_k)$ using the input signal $r(n)$ on which the perturbation is applied and the noisy output signals $y(n)$ and $u(n)$. Direct identification of $G(\Omega_k)$ through $y(n)$ and $u(n)$ leads to a biased estimate \Citep{Evers2020_FRF} since the two signals are correlated through the feedback loop by $v(n)$. 

Instead, an unbiased estimate of $G(\Omega_k)$ is obtained by identifying $\widehat{G}_{ru}$, given by 

\begin{equation}
   U(k) = \underbrace{\left(I+K(\Omega_k)G(\Omega_k)\right)^{-1}K(\Omega_k)}_{=\widehat{G}_{ru}(\Omega_k)} R(k)
\end{equation}

and $\widehat{G}_{ry}$, which is given by

\begin{equation}
   Y(k) = \underbrace{\left(I+G(\Omega_k)K(\Omega_k)\right)^{-1}G(\Omega_k)K(\Omega_k)}_{=\widehat{G}_{ry}(\Omega_k)} R(k),
\end{equation}

and then computing

\begin{equation}
   \widehat{G}(\Omega_k) = \widehat{G}_{ry}(\Omega_k) \widehat{G}_{ru}^{-1}(\Omega_k).
\end{equation}

The estimation of $\widehat{G}_{ry}$ and $\widehat{G}_{ru}$ are open-loop identification problems and the method presented in Section \ref{sec:FRF_using_LocalModels} can thus be used to obtain an estimate of $\widehat{G}_{ry}$ and $\widehat{G}_{ru}$.

\subsection{Uncertainty estimation}
\label{sec:UncertaintyEstimation}

To assess the quality of the estimated FRF, the variance of the estimate is approximated by computing the residuals of the local model estimate

\begin{equation}
   \widehat{V}(k) = Y_w(k)-\widehat{\Theta}(k)K_w(k).
   \label{eq:Residuals_Estimate}
\end{equation}

and the covariance of $G$ is given by \cite[eq.~(7.21)]{pintelon2012system}

\begin{equation}
   \mathrm{cov}(G(\Omega_k)) = S^H S \otimes \mathrm{cov}(\widehat{V}(k)), 
   \label{eq:Covariance_G_OL}
\end{equation}

with

\begin{equation}
   S = K_W^H(K_W K_W^H)^{-1} \begin{bmatrix} I_{n_u} \\ 0 \end{bmatrix}.
\end{equation}

To obtain the variance of $G$ using the indirect approach, the in-loop signals $Y(k)$ and $U(k)$ are combined into a single signal $Z(k) = \left[Y(k)\: U(k)\right]^T$ and the covariance of $G$ is then obtained by \Citep[eq.~(7.50)]{pintelon2012system} 

\begin{multline}
   \mathrm{cov}(G) = \\
   \left(\widehat{G}_{ru}^{-T}\otimes \left[I_{n_y} \: -\widehat{G}\right]\right)\mathrm{cov}(\widehat{G}_{rz})\left(\widehat{G}_{ru}^{-T}\otimes \left[I_{n_y} \: -\widehat{G}\right]\right)^H,
   \label{eq:Covariance_G_CL}
\end{multline}

where the frequency operator was left out for brevity purposes and the covariance matrix $\mathrm{cov}(\widehat{G}_{rz})$ is obtained from \eqref{eq:Covariance_G_OL}. The standard deviation of $G$ is then used to draw confidence bounds on the estimate of $G$. 

\subsection{Parameter selection}

The choice of the polynomial orders depends on the system dynamics. Based on the current experience with multiple sets of GW data, choosing the numerator orders $L_a=L_b=2$ typically leads to satisfactory results. For the denominator order, the choice depends on whether the system has lightly damped resonant dynamics, in which case a higher order up to $L_d=4$ is recommended. Without such dynamics, choosing $L_d=0$ is often sufficient but choosing a non-zero $L_d$ may yield marginally better results. For $W$, the total window size $2W+1$ needs to be larger than the number of parameters to be estimated. Any additional increase of $W$ will typically lead to better parameter estimates and thus a lower variance. However, the critical consideration is the bias-variance trade-off when selecting these parameters, as a higher $W$ will typically lead to a lower variance at the expense of an increased bias in the estimate. In practice, lower model orders like the ones proposed here and a window size with yields a few additional points to average over typically already lead to satisfactory results. Using this estimate as a baseline, then increasing the window size $W$ and comparing it to the original estimate can be helpful in determining when significant bias starts to occur.

\subsection{GraFIT}

GraFIT consists of a single function, written in both MATLAB and Python, performing FRF identification for systems operating in both open-loop and closed-loop and for arbitrary input and output dimensions. The example code here is all in MATLAB and the usage of the function is shown in Listing \ref{lst:GraFIT}.

\begin{lstlisting}[caption={GraFIT usage}, label={lst:GraFIT}]
[G,G_var,Gry,Gru,Gry_var,Gru_var,Y_contr] = ...
   GraFIT(r,y,u,W,freqIdentBand,Fs,L)
\end{lstlisting}

For systems operating in open-loop, the variable $r$ is the perturbation signal and $y$ the measured output, with the variable $u$ left blank. An example code for open-loop identification is shown in Listing \ref{lst:OLIdentificationCode}. Note here that the data must be stored as a three-dimensional matrix with the datapoints in the third dimension.

\begin{lstlisting}[caption={Basic open-loop identification}, label={lst:OLIdentificationCode}]
% Load data
% r = 1 x nr x N matrix of the input signals
% y = ny x nr x N matrix of the output signals

W = 12; % Size of window is 2W+1
freqIdentBand = [0.04 10]; % Frequencies to ...
                           % identify
Fs = 1e4; % Sampling frequency
La = 2; % Order of plant numerator
Lb = 2; % Order of transient numerator
Ld = 4; % Order of plant & transient denominator

[G,G_var] = ...
   GraFIT(r,y,[],W,freqIdentBand,Fs,[La Lb Ld]);
\end{lstlisting}

For systems operating in closed-loop, $r$ is again the perturbation signal and $u$ and $y$ are respectively chosen as the input and output of the dynamic system to be identified. An example code for indirect identification is shown in Listing \ref{lst:CLIdentificationCode}.

\begin{lstlisting}[caption={Basic indirect identification}, label={lst:CLIdentificationCode}]
% Load data
% r = 1 x nr x N matrix 
   % of the external perturbation
% y = ny x nr x N matrix 
   % of the output signal of the system 
% u = nu x nr x N matrix
   % of the input signal of the system

W = 18; % Size of window is 2W+1
freqIdentBand = [10 100]; % Frequencies to 
                          % identify
Fs = 1e4; % Sampling frequency
La = 2; % Order of plant numerator
Lb = 2; % Order of transient numerator
Ld = 2; % Order of plant & transient denominator

[G,G_var,Gry,Gru,Gry_var,Gru_var] = ...
   GraFIT(r,y,u,W,freqIdentBand,Fs,[La Lb Ld]);
\end{lstlisting}

\section{Experimental case studies}
\label{sec:ExperimentalCaseStudies}

In this section, the application of the toolbox on two of the three application domains from Section \ref{sec:ApplicationSetting} are presented to illustrate the effectiveness of the toolbox.

\subsection{Comparison with pre-existing approach: SA}

The baseline method for FRF measurements in GW detectors is the (SA) method. Consider again the standard identification problem as defined in Section \ref{sec:IdentificationProblem}, where a filterd white noise perturbation $r(n)$ is applied and the noisy output signal $y(n)$ is measured. The SA method obtains an estimate of $G(\Omega_k)$ by splitting the data into $P$ segments, multiplying it by a window (e.g., Von Hann) \Citep{Schoukens2006}, and computing

\begin{equation}
   G^{\mathrm{sa}}(\Omega_k) = \frac{\sum_{i=1}^{P}Y_i(k)\overline{R}_i(k)}{\sum_{i=1}^{P}R_i(k)\overline{R}_i(k)}.
   \label{eq:SAMethod}
\end{equation}s

The estimate of $G^{\mathrm{sa}}(\Omega_k)$ is improved for an increased number of segments $P$, which requires longer measurement times to obtain more segments. The SA method is the most common identification approach in GW detectors and is therefore used as a baseline for the comparison with the LRM method.

\subsection{Handling of noise, transient and leakage effects}

There are several effects which may affect the quality of the FRF. The first is the effect of the noise term $v(n)$, which is handled in both the LRM and SA method by averaging. In the LRM method, averaging is obtained by choosing the frequency window $W$ larger than the number of parameters to be estimated, while for the SA method, the data is split into $P$ segments and the estimate of $G^{\mathrm{sa}}(\Omega_k)$ is averaged over these segments. While both methods effectively handle the noise term, the downside of the SA method is that to obtain more segments, either longer datasets have to be measured or the data has to be split into segments which reduces the frequency resolution proportionaly by $P$. 

The second effect is dynamic transients in the system, represented by the term $T(\Omega_k)$. The SA method completely ignores this effect in the estimation, while the LRM method explicitly estimates this term and subsequently suppresses the effect of dynamic transients in the output, resulting in better estimates for systems with large dynamic transients \Citep{Evers2020_FRF}. The third effect is leakage, resulting from using non-periodic perturbation signals such as white noise. The SA method multiplies the segments by a window to mitigate the leakage effects, but it has been shown that this results in interpolation errors in the FRF \Citep{Gevers2011}. The LRM method handles leakage by also capturing this effect in the transient term $T(\Omega_k)$ and subsequently suppressing it in the output, which has been shown to be more effective in handling leakage effects \Citep{Gevers2011}.

\subsection{Assessment of estimation quality}
\label{sec:AssessmentEstimationQuality}

The coherence function $c\in\mathbb{R}_{\left[0,\: 1 \right]}$ \Citep[eq.~(2.47)]{pintelon2012system} is commonly used for the SA method to assess the FRF quality. The LRM method estimate provides standard deviation on the FRF estimate to assess quality, which can also be obtained from the coherence function \Citep[eq.~(7.14)]{pintelon2012system}. The main advantage of using the standard deviation is that it is much more interpretable, as it directly relates to the magnitude of the FRF. A clear advantage of the LRM method and toolbox in particular is that it also estimates the standard deviation of $G$ when the system is operating in closed-loop using \eqref{eq:Covariance_G_CL}. For the SA method, this is not directly possible since the covariances of $G_{ry}$ and $G_{ru}$ in \eqref{eq:Covariance_G_OL} are not directly available.

For the experimental case studies, the standard deviations on the estimates of the SA and LRM method are compared. To ensure a fair comparison, the number of segments used in the SA method is chosen to be equal to the number of the degrees of freedom $q$ of the LRM estimate, with $q$ given by \Citep[eq.~(7.14)]{pintelon2012system}

\begin{equation}
   q = 2W+1-n_{\Theta}.
   \label{eq:number_of_dofs_LRM}
\end{equation}

This essentially means the same number of averages for the estimation of the parameters are used for both methods.

\subsection{Direct identification for suspension system}

The first application domain is the suspension systems (see Section \ref{sec:appDomain_SuspensionSystems}) and the MultiSAS suspension \Citep{Heijningen2019} is used as an example system. This system is used to suspend the auxiliary optics of AdV+ such as photodiodes, for which the residual motion requirements are much less stringent, but it uses the same working principe as the suspensions for the main optics. Active control is applied on the first isolation stage of the system to damp the suspension modes and reduce the RMS motion. Three actuators and sensors are therefore located between the first isolation stage and the ground to damp the modes in the two horizontal translational directions ($x$, $z$) and the modes around the veritcal axes ($t_y$).

To assess the coupling between the Degrees of Freedom (DoF) and to design a controller for each DoF, the Multiple-Input Multiple-Output (MIMO) FRF of the system is measured in open-loop by consecutively injecting band-pass filtered white noise in each DoF. The system is identified using both the standard SA method and using the developed toolbox. For the SA method, $P=6$ segments (see \eqref{eq:SAMethod}) are used to have some averaging while maintaining enough frequency resolution. For the LRM method, a window size of $W=12$ is used, and the orders of the polynomials are chosen as $L_a=L_b=2$ and $L_d=4$ due to the largely undamped modes.

\begin{figure}[bt]
   \centering
   \resizebox{1.0\columnwidth}{!}{
           \includegraphics[]{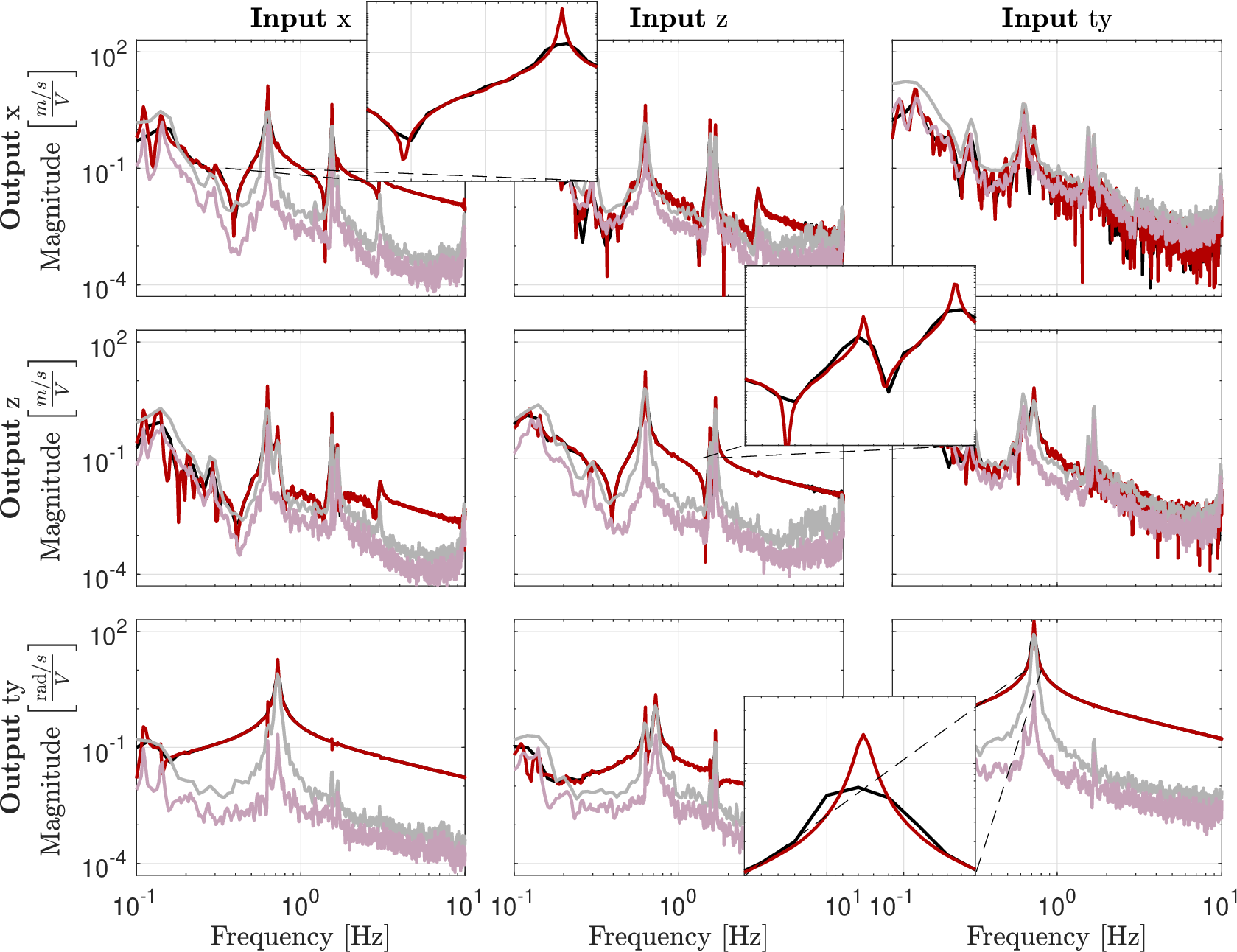}}
   \caption{FRF of $\widehat{G}$ for the $x$, $z$ and $t_y$ degrees of freedom of the top stage of the MultiSAS using SA estimation (\BlackLine) and its standard deviation (\GreyLine) and LRM estimation (\DarkRedLine) and its standard deviation (\lightRedPinkLine). The LRM method obtains almost consistently a factor 2 to 3 lower standard deviation with even lower standard deviations at the resonance peaks and also has a factor $P=6$ higher frequency resolution.}
   \label{fig:Suspension_LRM_vs_SA_F0}
\end{figure} 

The resulting estimates for both methods as well as their standard deviations are shown in Fig. \ref{fig:Suspension_LRM_vs_SA_F0}. The methods find almost identical estimates, indicating that the bias in both estimates is low, but the LRM method almost consistently achieves roughly a factor 2 lower standard deviation on the estimate with even larger standard deviation differences at the resonances. The reduced estimation error at the resonances provides a better estimate of the eigenfrequency and damping of the modes, which is desirable for the control design. An advantage of the LRM method is that the window size $W$ could be significantly increased to reduce the variance while maintaining the same frequency resolution, while for the SA method increasing the number of segments $P$ would reduce the frequency resolution by the proportional increase. However, the reduced variance for the LRM method will go at the expense of more bias so care has to be taken in not increasing $W$ too much.

\subsection{Indirect identification for optical system}

The second application domain is the longitudinal control of the mirrors in the AdV+ detector (see Section \ref{sec:appDomain_RelativeMirrorPositionControl}). The control system consists of 5 longitudinal DoFs but only the DARM, MICH and SRCL DoFs \Citep{Bersanetti_AdV_Status} will be considered here for clarity of presentation. The control loop uses error signals derived from the photodiodes to actuate on the mirror positions. To identify the FRF, the system must operate in closed-loop and bandpass filtered white noise is consecutively injected in each DoF for \SI{120}{\second}. For the SA method, $P=24$ segments are used and a Von Hann window is applied to deal with the significant leakage occuring. For the LRM method, a window size of $W=18$ is used and the orders of the polynomials are chosen as $L_a=L_b=L_d=2$. The indirect identification approach from Section \ref{sec:IndirectIdentification} is used to obtain an estimate of $G$. 

\begin{figure}[bt]
   \centering
   \resizebox{1.0\columnwidth}{!}{
           \includegraphics[]{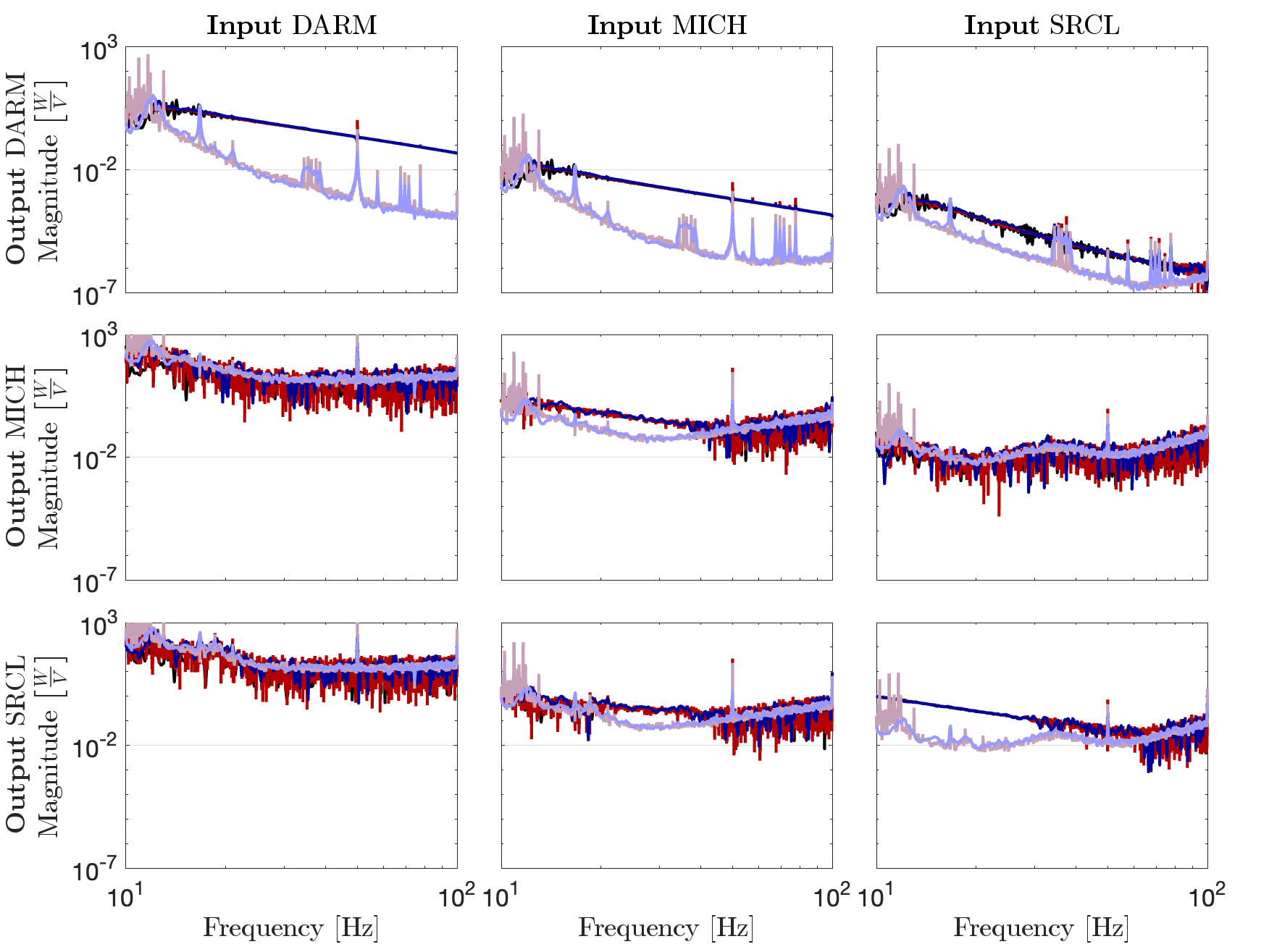}}
   \caption{FRF of $\widehat{G}$ for the three longitudinal DoFs using SA estimation (\BlackLine) and its standard deviation (\GreyLine) for \SI{120}{\second} of data, LRM estimation (\DarkRedLine) and its standard deviation (\lightRedPinkLine) for \SI{120}{\second} of data and LRM estimation (\DarkBlueLine) and its standard deviation (\lightBlueLine) for the first  \SI{12}{\second} of the same dataset. The standard deviation for SA is not available. Just \SI{12}{\second} of data using the LRM method is sufficient to get a good quality FRF.}
   \label{fig:LSC_Plant_withUncertainty}
\end{figure}

In Fig. \ref{fig:LSC_Plant_withUncertainty}, the estimate of $G$ for the SA method with \SI{120}{\second} of data and the LRM method with \SI{120}{\second} and \SI{12}{\second} of data are shown. The standard deviations for both LRM methods are also shown, but not for the SA method since it is not directly available (see Section \ref{sec:AssessmentEstimationQuality}). All three estimates are roughly identical, indicating that the bias is likely low for all three estimates. The LRM method with just \SI{12}{\second} of data furthermore has roughly equal standard deviations as the LRM method with \SI{120}{\second} of data, illustrating that the LRM method is very effective in using minimal data for the estimation.

\begin{figure}[bt]
   \centering
   \resizebox{1.0\columnwidth}{!}{
           \includegraphics[]{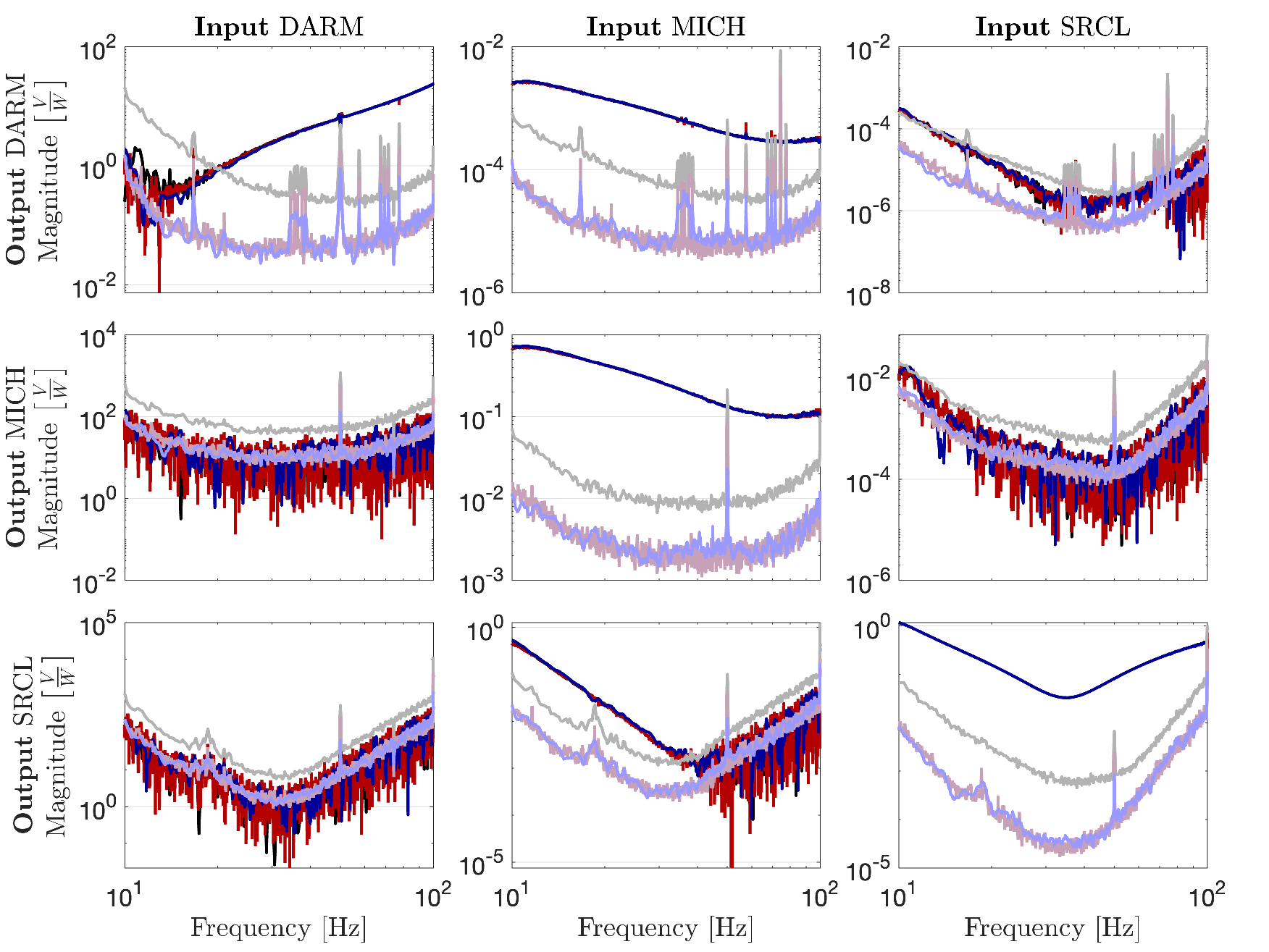}}
   \caption{FRF of $\widehat{G}_{ru}$ for the three longitudinal DoFs using SA estimation (\BlackLine) and its standard deviation (\GreyLine) for \SI{120}{\second} of data, LRM estimation (\DarkRedLine) and its standard deviation (\lightRedPinkLine) for \SI{120}{\second} of data and LRM estimation (\DarkBlueLine) and its standard deviation (\lightBlueLine) for the first \SI{12}{\second} of the same dataset. The LRM method is able to obtain an order lower variance with just a tenth of the data compared to SA, while having almost identical estimates, indicating marginal bias.}
   \label{fig:LSC_CS_measTimeComp}
\end{figure}

To have a more direct assessment of the LRM versus SA method, the estimate of $G_{ru}$ is shown in Fig. \ref{fig:LSC_CS_measTimeComp} (similair observations for $G_{ry}$ have been observed and this plot has therefore been left out for brevity), together with the standard deviations for the three estimates. The LRM method using both \SI{120}{\second} and \SI{12}{\second} of data obtains almost consistently an order lower standard deviation compared to the SA method with \SI{120}{\second} of data, while the bias is almost identical for all three estimates. The LRM method thus obtains a factor 10 lower standard deviations compared to the SA method using the same data for the estimation. Furthermore, it is also much more effective with less data, obtaining significantly lower standard deviations with 10 times less data.

\begin{figure}[bt]
   \centering
   \resizebox{1.0\columnwidth}{!}{
           \includegraphics[]{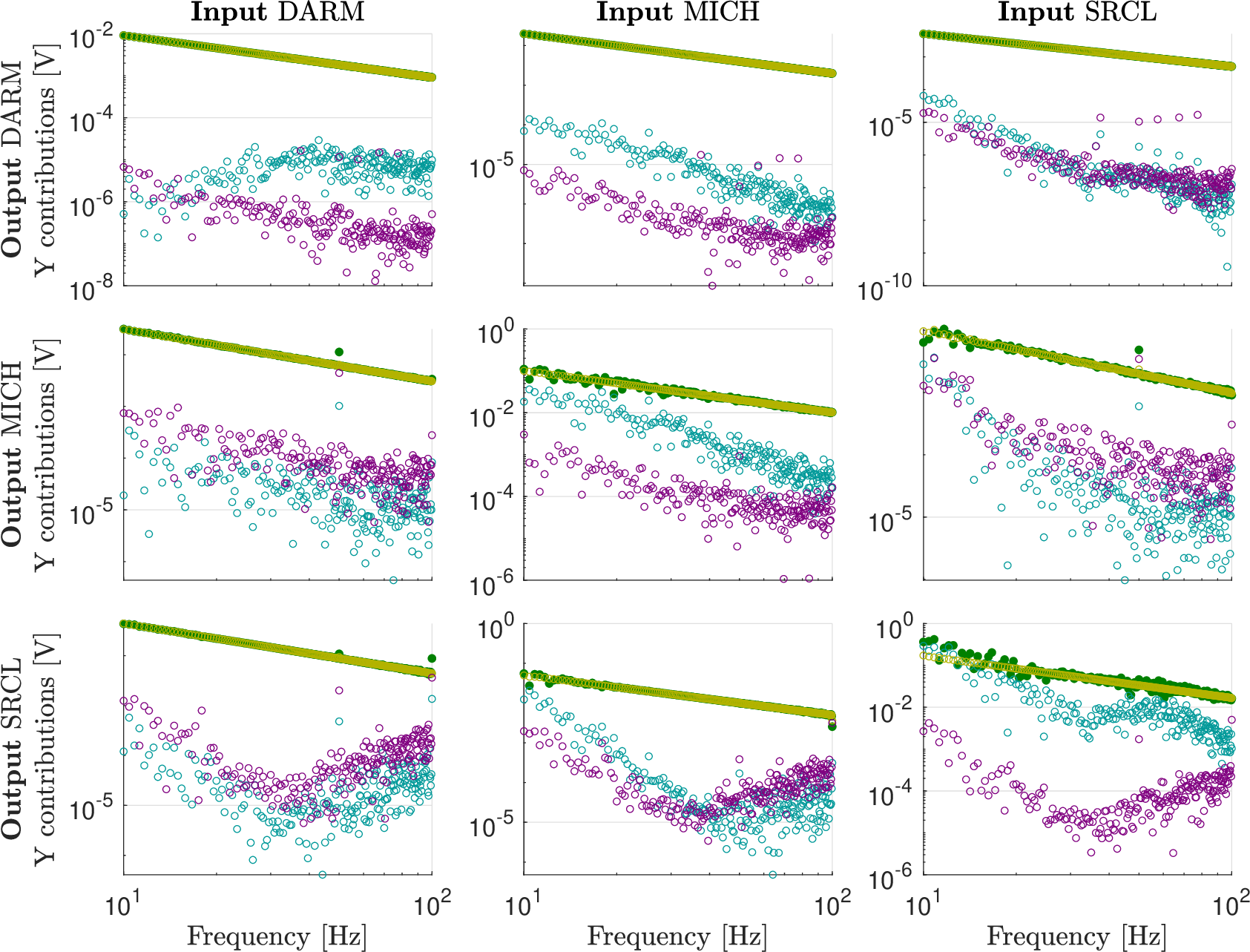}}
   \caption{Output DFT $U(k)$ (\darkGreenCircle), input contribution $G_{ru}(\Omega_k)R(k)$ (\darkCyanCircle), transient contribution $T(\Omega_k)$ (\goldCircle) and noise $V(k)$ (\purpleCircle). The DFTs are downsampled in the frequency domain for clarity of presentation. The transient contribution dominates the output DFT for almost all entries.}
   \label{fig:LSC_Gru_Output_contributions}
\end{figure}

One of the reasons for the substantially better estimate using LRM is the handling of leakage effects, which are significant in the DFT of $U(k)$. In Fig. \ref{fig:LSC_Gru_Output_contributions}, the contributions of the input, transient and noise term to the DFT of $u$ are shown, which have all been estimated using the LRM method. The contribution of the transient term, which identifies the leakage effect, is in almost every entry at least an order higher than the input contribution. By directly estimating the transient effect in the LRM method, rather than using windowing as in the SA method, the leakage effects are substantially reduced, leading to a better estimate of $G_{ru}$ and therefore $G$.

\section{Conclusions}
\label{sec:Conclusions}
The two main requirements for identifying FRFs is that they are accurate and use minimal data. Applying GraFIT to two application domains in GW detectors has shown that the LRM method in the toolbox achieves, depending on the application domain, from at least a factor 2 to more than order of magnitude lower standard deviations compared to the SA method. The LRM method has furthermore been shown to achieve better quality FRFs with 10 times less data than the SA method, thus providing more accurate FRFs with less data than the SA method. Additionally, GraFIT handles systems operating in closed-loop as well as systems with arbitrary input-output dimensions, significantly simplifying the identification procedure, with the only downside being increased computation time by a few factors. This makes GraFIT a valuable tool for the identification of FRFs in Gravitational Wave detectors, where minimal data use is essential.

\begin{ack}
The authors gratefully acknowledge the contributions made by Luuk van Vliet for his testing on GW data. The authors also would like to acknowledge the contributions made by Rick van der Maas, Annemiek van Rietschoten, Enzo Evers, Robbert Voorhoeve and Paul Tacx in developing the toolbox. The authors also gratefully acknowledge the contributions made by the ISC team at Advanced Virgo+ for helping to perform the experiments on the interferometer as well as providing the necessary information and support to conduct this work. The authors also gratefully acknowledge the Italian Istituto Nazionale di Fisica Nucleare (INFN), the French Centre National de la Recherche Scientifique (CNRS) and the Netherlands Organization for Scientific Research, for the construction and operation of the Virgo detector and the creation and support of the EGO consortium. The authors also gratefully acknowledge research support from these agencies as well as by  the Spanish  Agencia Estatal de Investigaci\'on, the Consellera d'Innovaci\'o, Universitats, Ci\`encia i Societat Digital de la Generalitat Valenciana and the CERCA Programme Generalitat de Catalunya, Spain, the National Science Centre of Poland and the Foundation for Polish Science (FNP), the European Commission, the Hungarian Scientific Research Fund (OTKA), the French Lyon Institute of Origins (LIO), the Belgian Fonds de la Recherche Scientifique (FRS-FNRS), Actions de Recherche Concertées (ARC) and Fonds Wetenschappelijk Onderzoek – Vlaanderen (FWO), Belgium.

\end{ack}

\bibliography{ifacconf}             

\begin{thebibliography}{20}
\providecommand{\natexlab}[1]{#1}
\providecommand{\url}[1]{\texttt{#1}}
\providecommand{\urlprefix}{URL }
\expandafter\ifx\csname urlstyle\endcsname\relax
  \providecommand{\doi}[1]{doi:\discretionary{}{}{}#1}\else
  \providecommand{\doi}{doi:\discretionary{}{}{}\begingroup
  \urlstyle{rm}\Url}\fi

\bibitem[{Acernese et~al.(2015)}]{Acernese_Virgo_Overview}
Acernese, F. et~al. (2015).
\newblock {Advanced Virgo: a second-generation interferometric gravitational
  wave detector}.
\newblock \emph{Class. Quant. Grav.}, 32(2), 024001.

\bibitem[{Bendat and Piersol(1980)}]{Bendat_SA}
Bendat, J.S. and Piersol, A.G. (1980).
\newblock \emph{Engineering Applications of Correlation and Spectral Analysis}.
\newblock Wiley, New York.

\bibitem[{Bersanetti et~al.(2022)Bersanetti, Boldrini, Diaz, Freise, Maggiore,
  Mantovani, and Valentini}]{Bersanetti_DRMI_ErrorSignals}
Bersanetti, D., Boldrini, M., Diaz, J.C., Freise, A., Maggiore, R., Mantovani,
  M., and Valentini, M. (2022).
\newblock {Simulations for the Locking and Alignment Strategy of the DRMI
  Configuration of the Advanced Virgo Plus Detector}.
\newblock \emph{Galaxies}, 10(6).

\bibitem[{Bersanetti et~al.(2021)Bersanetti, Patricelli, Piccinni,
  Piergiovanni, Salemi, and Sequino}]{Bersanetti_AdV_Status}
Bersanetti, D., Patricelli, B., Piccinni, O.J., Piergiovanni, F., Salemi, F.,
  and Sequino, V. (2021).
\newblock {Advanced Virgo: Status of the Detector, Latest Results and Future
  Prospects}.
\newblock \emph{Universe}, 7(9).

\bibitem[{Braccini et~al.(2005)}]{braccini_SuperAttenuator_Perf_Virgo}
Braccini, S. et~al. (2005).
\newblock {Measurement of the seismic attenuation performance of the VIRGO
  Superattenuator}.
\newblock \emph{Astroparticle Physics}, 23(6), 557--565.

\bibitem[{Buikema et~al.(2020)}]{Buikema_aLIGO_StatusO3}
Buikema, A. et~al. (2020).
\newblock {Sensitivity and performance of the Advanced LIGO detectors in the
  third observing run}.
\newblock \emph{Phys. Rev. D}, 102, 062003.

\bibitem[{Evers et~al.(2020{\natexlab{a}})Evers, {van Tuijl}, Lamers, {de
  Jager}, and Oomen}]{Evers2020_ThermalIdentification}
Evers, E., {van Tuijl}, N., Lamers, R., {de Jager}, B., and Oomen, T.
  (2020{\natexlab{a}}).
\newblock Fast and accurate identification of thermal dynamics for precision
  motion control: Exploiting transient data and additional disturbance inputs.
\newblock \emph{Mechatronics}, 70, 102401.

\bibitem[{Evers et~al.(2020{\natexlab{b}})Evers, Voorhoeve, and
  Oomen}]{Evers2020_FRF}
Evers, E., Voorhoeve, R., and Oomen, T. (2020{\natexlab{b}}).
\newblock On frequency response function identification for advanced motion
  control.
\newblock In \emph{2020 IEEE 16th International Workshop on Advanced Motion
  Control (AMC)}, 1--6.

\bibitem[{Gevers et~al.(2011)Gevers, Pintelon, and Schoukens}]{Gevers2011}
Gevers, M., Pintelon, R., and Schoukens, J. (2011).
\newblock The local polynomial method for nonparametric system identification:
  Improvements and experimentation.
\newblock In \emph{2011 50th IEEE Conference on Decision and Control and
  European Control Conference}, 4302--4307.

\bibitem[{H{\"a}gg et~al.(2016)H{\"a}gg, Schoukens, Gevers, and
  Hjalmarsson}]{Hagg2016}
H{\"a}gg, P., Schoukens, J., Gevers, M., and Hjalmarsson, H. (2016).
\newblock The transient impulse response modeling method for non-parametric
  system identification.
\newblock \emph{Automatica}, 68, 314--328.

\bibitem[{Heijningen et~al.(2019)Heijningen, Bertolini, Hennes, Beker, Doets,
  Bulten, Agatsuma, Sekiguchi, and van~den Brand}]{Heijningen2019}
Heijningen, J., Bertolini, A., Hennes, E., Beker, M., Doets, M., Bulten, H.,
  Agatsuma, K., Sekiguchi, T., and van~den Brand, J. (2019).
\newblock A multistage vibration isolation system for advanced virgo suspended
  optical benches.
\newblock \emph{Classical and Quantum Gravity}, 36.

\bibitem[{Lataire and Chen(2016)}]{Lataire2016}
Lataire, J. and Chen, T. (2016).
\newblock Transfer function and transient estimation by gaussian process
  regression in the frequency domain.
\newblock \emph{Automatica}, 72, 217--229.

\bibitem[{McKelvey and Gu{\'e}rin(2012)}]{McKelvey2012}
McKelvey, T. and Gu{\'e}rin, G. (2012).
\newblock Non-parametric frequency response estimation using a local rational
  model1.
\newblock \emph{IFAC Proceedings Volumes}, 45(16), 49--54.
\newblock 16th IFAC Symposium on System Identification.

\bibitem[{Pintelon and Schoukens(2012)}]{pintelon2012system}
Pintelon, R. and Schoukens, J. (2012).
\newblock \emph{System Identification: A Frequency Domain Approach (2nd ed.)}.
\newblock John Wiley.

\bibitem[{Schoukens et~al.(2005)Schoukens, Pintelon, Dobrowiecki, and
  Rolain}]{Schoukens2005}
Schoukens, J., Pintelon, R., Dobrowiecki, T., and Rolain, Y. (2005).
\newblock Identification of linear systems with nonlinear distortions.
\newblock \emph{Automatica}, 41(3), 491--504.
\newblock Data-Based Modelling and System Identification.

\bibitem[{Schoukens et~al.(2006)Schoukens, Rolain, and
  Pintelon}]{Schoukens2006}
Schoukens, J., Rolain, Y., and Pintelon, R. (2006).
\newblock Analysis of windowing/leakage effects in frequency response function
  measurements.
\newblock \emph{Automatica}, 42(1), 27--38.

\bibitem[{Schoukens et~al.(2009)Schoukens, Vandersteen, Barb{\'e}, and
  Pintelon}]{Schoukens2009}
Schoukens, J., Vandersteen, G., Barb{\'e}, K., and Pintelon, R. (2009).
\newblock Nonparametric preprocessing in system identification: A powerful
  tool.
\newblock In \emph{2009 European Control Conference (ECC)}, 1--14.

\bibitem[{Tacx et~al.(2024)Tacx, Habraken, Witvoet, Heertjes, and
  Oomen}]{Tacx2024}
Tacx, P., Habraken, R., Witvoet, G., Heertjes, M., and Oomen, T. (2024).
\newblock Identification of an overactuated deformable mirror system with
  unmeasured outputs.
\newblock \emph{Mechatronics}, 99, 103158.

\bibitem[{van Dael et~al.(2024)}]{vanDael2024_OnlineDecoupling}
van Dael, M. et~al. (2024).
\newblock {Online decoupling of the time-varying longitudinal feedback loops
  for improved performance in Advanced Virgo Plus$^{*}$}.
\newblock \emph{Class. Quant. Grav.}, 41(21), 215008.

\bibitem[{Voorhoeve et~al.(2018)Voorhoeve, {van der Maas}, and
  Oomen}]{VOORHOEVE2018129}
Voorhoeve, R., {van der Maas}, A., and Oomen, T. (2018).
\newblock Non-parametric identification of multivariable systems: A local
  rational modeling approach with application to a vibration isolation
  benchmark.
\newblock \emph{Mechanical Systems and Signal Processing}, 105, 129--152.

\end{thebibliography}
                                                   
\end{document}